\documentclass[12pt]{article}
\usepackage{epsfig,graphicx}
\usepackage{subfigure}
\usepackage{times}

\title{Anomalous Strength Characteristics of Tilt Grain Boundaries in Graphene}

 \author
{Rassin Grantab$^{1}$, Vivek B. Shenoy$^{1\ast}$, Rodney S. Ruoff$^{2\ast}$\\
\\
\normalsize{$^{1}$Division of Engineering, Brown University, Providence, RI, 02906, USA}\\
\qquad\\
\normalsize{$^{2}$Department of Mechanical Engineering and The Texas Materials Institute,}\\
\normalsize{The University of Texas, Austin, TX, 78712, USA}\\
\qquad\\
\normalsize{$^\ast$To whom correspondence should be addressed: vivek\_shenoy@brown.edu; r.ruoff@mail.utexas.edu}\\
}

\date{}

\begin{document} 
\setlength\fboxsep{0pt}
\baselineskip24pt

\maketitle

\begin{abstract}

Using molecular dynamics simulations and first principles calculations, we have studied the structure and mechanical strength of tilt grain boundaries in graphene sheets that arise during CVD growth of graphene on metal substrates. Surprisingly, we find that for tilt boundaries in the vicinity of both the zig-zag and arm-chair orientations, large angle boundaries with a higher density of 5-7 defect pairs are \emph{stronger} than the low-angle boundaries which are comprised of fewer defects per unit length. Interestingly, the trends in our results cannot be explained by a continuum Griffith-type fracture mechanics criterion, which predicts the opposite trend due to that fact that it does not account for the critical bonds that are responsible for the failure mechanism. We have identified the highly-strained bonds in the 7-member rings that lead to the failure of the sheets, and we have found that large angle boundaries are able to better accommodate the strained 7-rings. Our results provide guidelines for designing growth methods to obtain grain boundary structures that can have strengths close to that of pristine graphene.

\end{abstract}

\newpage

Graphene continues to be one of the most widely researched areas in materials science today. It is the thinnest material ever synthesized, yet the one of the strongest ever measured \cite{geim_2009, lee_2008}, and it 
exhibits exceptional electronic, thermal, and optical properties \cite{geim_2009, geim_2007}. However, despite the prolific research efforts, growing large-area, single-layer graphene sheets remains a challenge. Recently, a chemical vapor deposition (CVD) technique has been devised that exploits the low solubility of carbon in metals such as nickel \cite{yu_2008,kim_2009} and copper \cite{li_2009, levendorf_2009} in order to grow graphene on metal foils. A consequence of this technique is that the large-area graphene sheets contain grain boundaries, since each grain in the metallic foil serves as a nucleation site for individual grains of graphene \cite{li_2009}.

Tilt grain boundaries in graphite had been observed in scanning tunneling microscopy (STM) experiments over 20 years ago by Albrecht \emph{et al.} \cite{albrecht_1988}, and since then several groups have performed similar microscopy studies \cite{osing_1998,simonis_2002,pong_2007,cervenka_2009,wong_2009}. More recently, Hashimoto \emph{et al.} \cite{hashimoto_2004} have observed individual dislocations in graphene using transmission electron microscopy (TEM), and the structure, as well as the electronic, magnetic, and dynamical properties of grain boundaries in graphene have been investigated by a number of other research teams \cite{yazyev_2010, malola_2010,cervenka_2009b}. With all this previous work established, a natural question to ask is: how do these grain boundaries influence the mechanical properties of graphene? Given the fact that graphene is one of the stiffest (modulus ${\sim 1TPa}$) and strongest materials (strength ${\sim 100GPa}$), in order to use CVD-synthesized graphene sheets in NEMS, sensors, and as pressure barriers, it is important to know how the grain boundaries influence these fundamental mechanical properties.

Although a number of studies have been carried out on the mechanics of dislocations and defects in carbon nanotubes \cite{li_2005,huang_2008, zhang_2005} and graphene \cite{zhang_2006}, the mechanical properties of hydrogen-functionalized graphene \cite{pei_2010}, as well as the fracture and failure of graphene and carbon nanotubes with multiple vacancies \cite{xiao_2009} and Stone-Wales defects \cite{xiao_2009,xiao_2010,lu_2005}, the effect of grain boundaries on the mechanical properties of graphene has been largely neglected. In this paper, we address this outstanding problem using Molecular Dynamics and first-principles calculations. Contrary to the intuitive picture based on continuum models that more defective structures lead to greater deterioration of mechanical properties, we find that as the grain boundary angles, and hence the number of defects per unit-length, increase, the strengths of the graphene sheets also increase. We have identified the underlying reason for this counterintuitive phenomenon by analyzing the initial strains within the carbon-carbon bonds along the grain boundaries. Our first-principles calculations show that higher-angle tilt grain boundaries are able to better accommodate the strained 7-member rings, which explains the increased strength. Our results indicate that CVD-grown 'polycrystalline' sheets can be comparable in strength to pristine graphene provided that they are comprised largely of higher-angle tilt grain boundaries.


Molecular Dynamics (MD) simulations were performed using the MD package, LAMMPS \cite{plimpton_1995}. All simulations were performed on ${80}${\AA} square graphene sheets; the tilt grain boundaries were placed in the middle of the sheets and were oriented parallel to the Y-direction. The graphene sheets were deformed under tensile loading in directions perpendicular (along the X-axis) and parallel (along the Y-axis) to the grain boundaries at a constant strain-rate until complete failure was observed. A ${2.5}${\AA} wide strip of material at each end of the sheet was constrained against motion along the direction of deformation (but free to move in the direction perpendicular to the direction of deformation) by enforcing zero force and velocity on the atoms in these regions. With these constraints in place, the sheets were subsequently relaxed for ${10000}$ MD steps, then a homogeneous strain of ${0.5\%}$ was applied to the graphene sheets by scaling all atomic coordinates accordingly; using a time step of ${1fs}$, this results in an average strain-rate of ${0.05\%ps^{-1}}$. This procedure of relaxation and stretching was applied sequentially until complete failure of each graphene sheet. All MD simulations were performed using an NVE ensemble. 

An adaptive intermolecular reactive bond order (AIREBO) potential \cite{stuart_2000} as implemented in LAMMPS, was used to model the atomic interactions in graphene. Following the work of Pei \emph{et al.} \cite{pei_2010}, we have used an interaction cut-off parameter of ${1.92}${\AA}. In order to calculate the stress-strain curves during deformation, the stress on each individual carbon atom was first calculated according to the following Virial stress expression \cite{basinski_1971,chandra_2004}:

\begin{equation}
\sigma_{ij}^{\alpha}=\frac{1}{\Omega^{\alpha}}\Bigg(\frac{1}{2}m^{\alpha}v_i^{\alpha}v_j^{\alpha}+\sum_{\beta=1,n} r_{\alpha \beta}^j f_{\alpha \beta}^i\Bigg)
\label{eq1}
\end{equation}

In the equation above, ${i}$ and ${j}$ denote the indices in the Cartesian coordinate system, while ${\alpha}$ and ${\beta}$ are the atomic indices; ${m^{\alpha}}$ and ${v^{\alpha}}$ are the mass and velocity of atom ${\alpha}$, respectively; ${r_{\alpha \beta}}$ and ${f_{\alpha \beta}}$ are the distance and force between atoms ${\alpha}$ and ${\beta}$, respectively; and ${\Omega^{\alpha}}$ is the atomic volume of atom ${\alpha}$. Once the stress on each atom was computed, we then averaged the stress over the entire sheet every ${500}$ MD time-steps, and averaged these values over the latter half of the relaxation period of ${10000}$ time-steps in order to obtain both a spatial and temporal average of the stresses. This method provides a single stress value for every strain increment, thereby allowing us to construct a stress-strain curve for the graphene sheets.

The inter-atomic potential and  simulation method as a whole were validated by deforming pristine graphene and comparing the results to experiments. Our methods predicted an elastic modulus of ${0.8TPa}$, an ultimate strength of ${125GPa}$, and a strain-at-failure of ${25\%}$ for zig-zag oriented graphene. Our predicted value of elastic modulus is within ${20\%}$ of the experimental value reported by Lee \emph{et al.} \cite{lee_2008}, while the ultimate strength and strain at failure match the experimental values almost exactly.

The first-principles Density Functional Theory (DFT) calculations were performed with a plane-wave basis-set using the \emph{ab initio} simulation package, VASP \cite{kresse_1996,kresse_1996b}. Projector-Augmented Wave potentials (PAW) \cite{kresse_1999} were used to represent the ionic cores, and Perdew-Burke-Ernzerhof (PBE) exchange-correlation functionals \cite{perdew_1996} were used for gradient approximations. For all the DFT calculations, a vacuum of ${12}${\AA} was used in the direction perpendicular to the graphene sheets, and the sheets were periodic in the direction parallel to the grain boundaries (the Y-direction). In the X-direction, we saturated the non-periodic graphene edges with hydrogen atoms in order to ensure that all carbon atoms were ${sp^2}$ bonded (the Z-direction is normal to the graphene basal-plane). A kinetic energy cutoff of ${500}$eV was used in all DFT calculations. The structures were relaxed using the conjugate-gradient algorithm until the atomic forces were smaller than ${0.04}${eV}/{\AA}. A convergence study was performed in which the k-point mesh was varied from ${1\times5\times1}$ up to ${1\times20\times1}$, with the key results varying by no more than ${0.2\%}$  (only 1 k-point has been used in the Y- and Z-directions since they are both non-periodic). A separate study was also carried out in which the width of the model (in the X-direction) was varied from ${18.7}${\AA} to ${28.9}${\AA}; in this case, the maximum discrepancy in the results was less than ${1\%}$.

\begin{figure}[!ht]
    \centering
    \subfigure[${5.5^{\circ}}$]
    { \label(${5.5^{\circ}}$){fig1_sub1}
        \fbox{\includegraphics[scale=1.00]{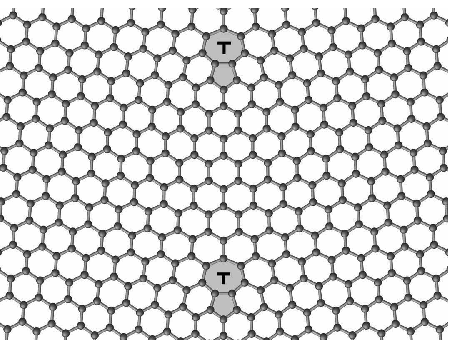}}
    } \qquad
    \subfigure[${13.2^{\circ}}$]
    { \label(${13.2^{\circ}}$){fig1_sub2}
        \fbox{\includegraphics[scale=1.00]{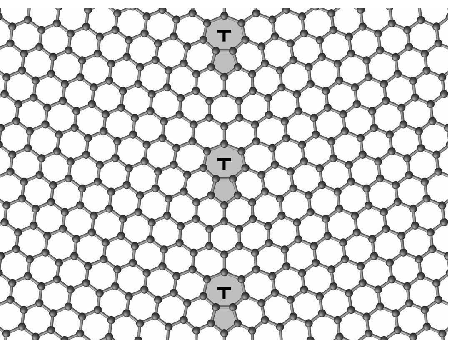}}
    } \qquad
    \subfigure[${21.7^{\circ}}$]
    { \label(${21.7^{\circ}}$){fig1_sub3}
       \fbox{\includegraphics[scale=1.00]{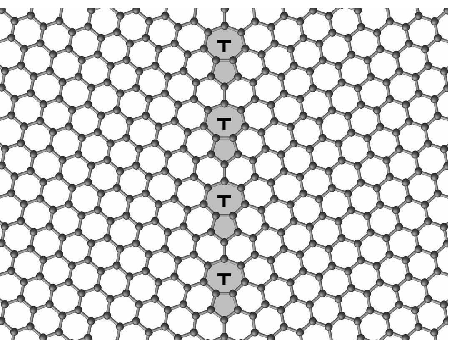}}
    } \\
    \caption{The structures of grain boundaries in zig-zag oriented graphene sheets with varying mismatch angles.}
    \label{fig1}
\end{figure}

The structures of tilt-grain boundaries in zig-zag oriented graphene are shown in Figure \ref{fig1} for grain boundary angles of ${5.5^{\circ}}$, ${13.2^{\circ}}$, and ${21.7^{\circ}}$ (the angles represent the total mismatch angles between the left and right grains). The grain boundaries consist of repeating 5-7 ring pairs that are separated by several hex-rings. As the grain boundary angle increases, the number of hex-rings separating the 5-7 defects decreases, with the ultimate limit occurring at ${21.7^{\circ}}$ when only a single hex-ring separates the periodic 5-7 defects. Therefore, more severe  grain boundary angles are comprised of higher defect densities. The repeating defect pairs can also be thought of as an array of edge dislocations with horizontal Burgers vectors where the 5-rings represent the extra plane of atoms, as shown in Figure \ref{fig1}.

\begin{figure}[!ht]
    \centering
    \subfigure[${15.8^{\circ}}$]
    { \label(${15.8^{\circ}}$){fig2_sub1}
        \fbox{\includegraphics[scale=1.00]{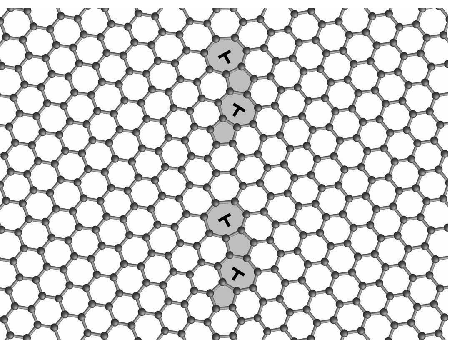}}
    } \qquad
    \subfigure[${21.4^{\circ}}$]
    { \label(${21.4^{\circ}}$){fig2_sub2}
        \fbox{\includegraphics[scale=1.00]{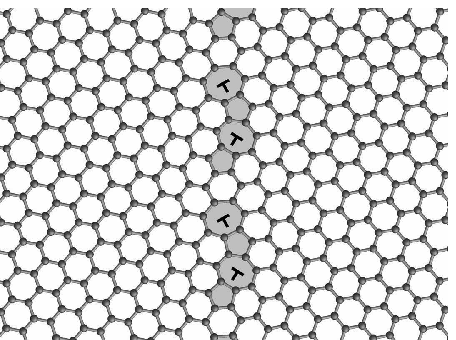}}
    } \qquad
    \subfigure[${28.7^{\circ}}$]
    { \label(${28.7^{\circ}}$){fig2_sub3}
        \fbox{\includegraphics[scale=1.00]{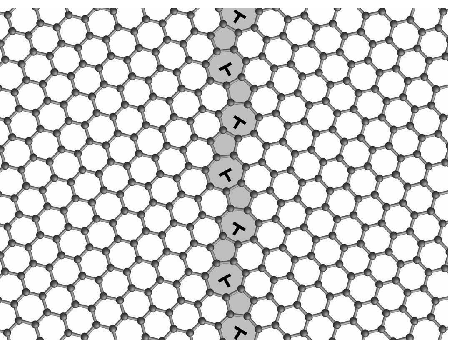}}
    } \\
    \caption{The structures of grain boundaries in arm-chair oriented graphene sheets with varying mismatch angles.}
    \label{fig2}
\end{figure}

Tilt boundaries in arm-chair oriented graphene are shown in Figure \ref{fig2} for grain boundary angles of ${15.8^{\circ}}$, ${21.4^{\circ}}$, and ${28.7^{\circ}}$. For this orientation, the repeating defect consists of two diagonally opposed 5-7 pairs that are separated by several hex rings. As was the case for the zig-zag orientation, larger grain boundary angles consist of higher defect densities; however, for the arm-chair oriented graphene, the most severely misoriented boundary (${28.7^{\circ}}$) consists of repeating 5-7 pairs without any intermediate hex-rings. Viewing the grain boundary in terms of dislocations, the two diagonally opposed, repeating 5-7 pairs represent two partial edge dislocations, as shown in Figure \ref{fig2}. The vertical components of the Burgers vectors of the two partial dislocations nullify one-another, leaving the grain boundary vertically oriented (for a vertical boundary, the net Burgers vector must be purely horizontal).

\begin{figure}[!ht]
    \centering
    \subfigure[Perpendicular]
    { \label([Perpendicular]){fig3_sub1}
        \includegraphics[scale=0.5]{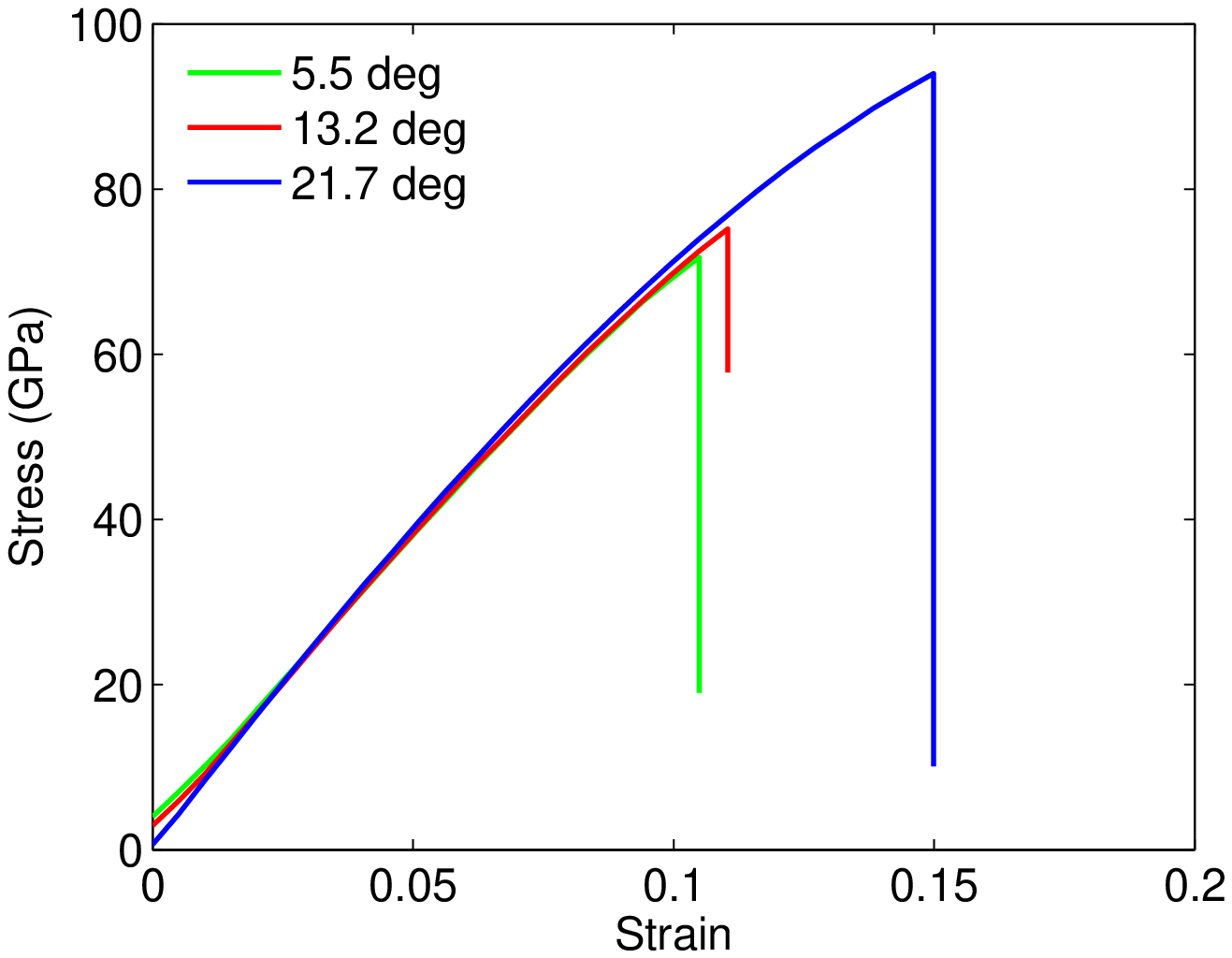}
    } \qquad
    \subfigure[Parallel]
    { \label(Parallel){fig3_sub2}
        \includegraphics[scale=0.5]{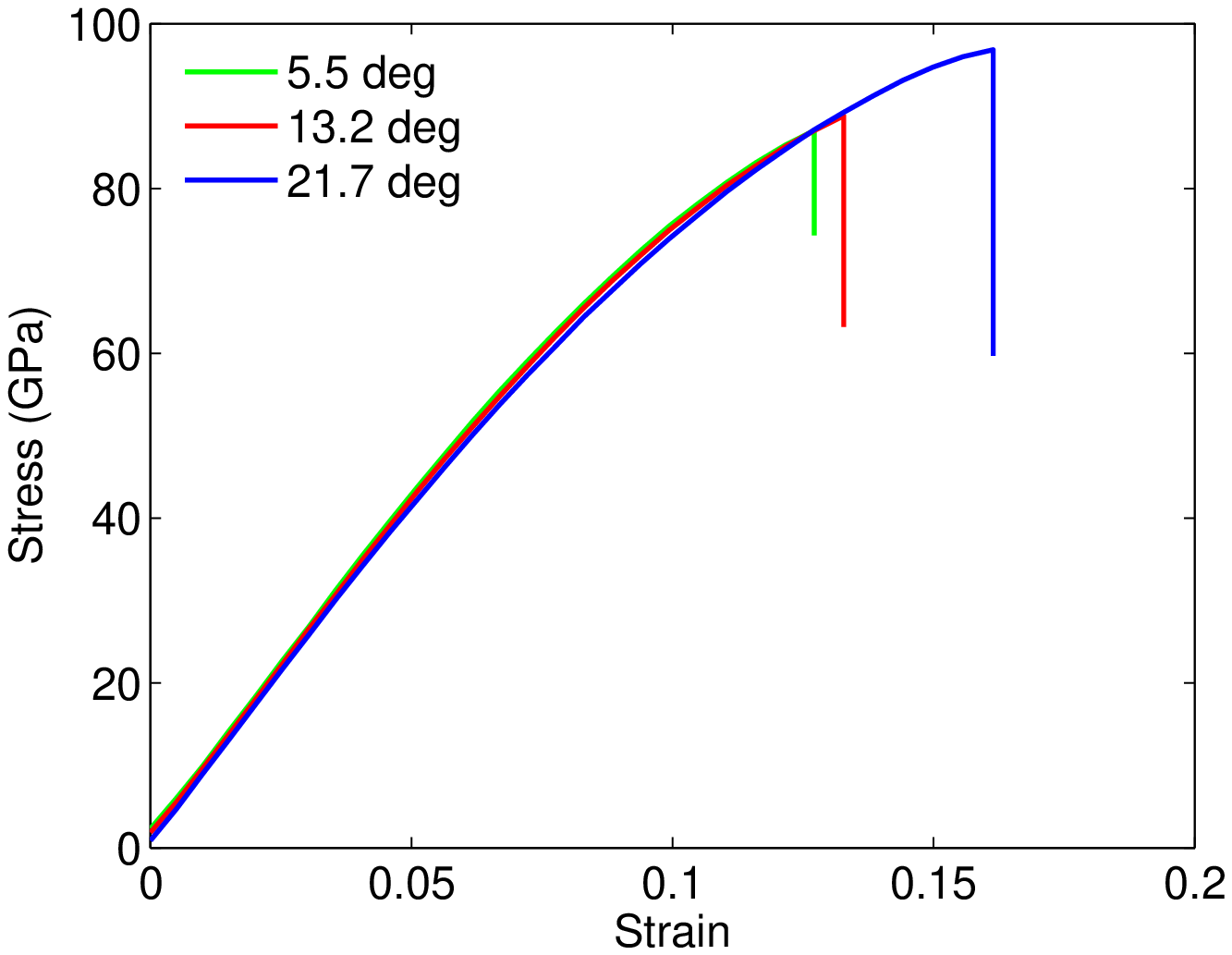} 
    } \\
    \caption{The stress-strain curves of zig-zag oriented graphene sheets pulled perpendicular (a) and parallel (b) to the grain boundaries.}
    \label{fig3}
\end{figure}

\begin{figure}[!ht]
    \centering
    \subfigure[Perpendicular]
    { \label([Perpendicular]){fig4_sub1}
        \includegraphics[scale=0.5]{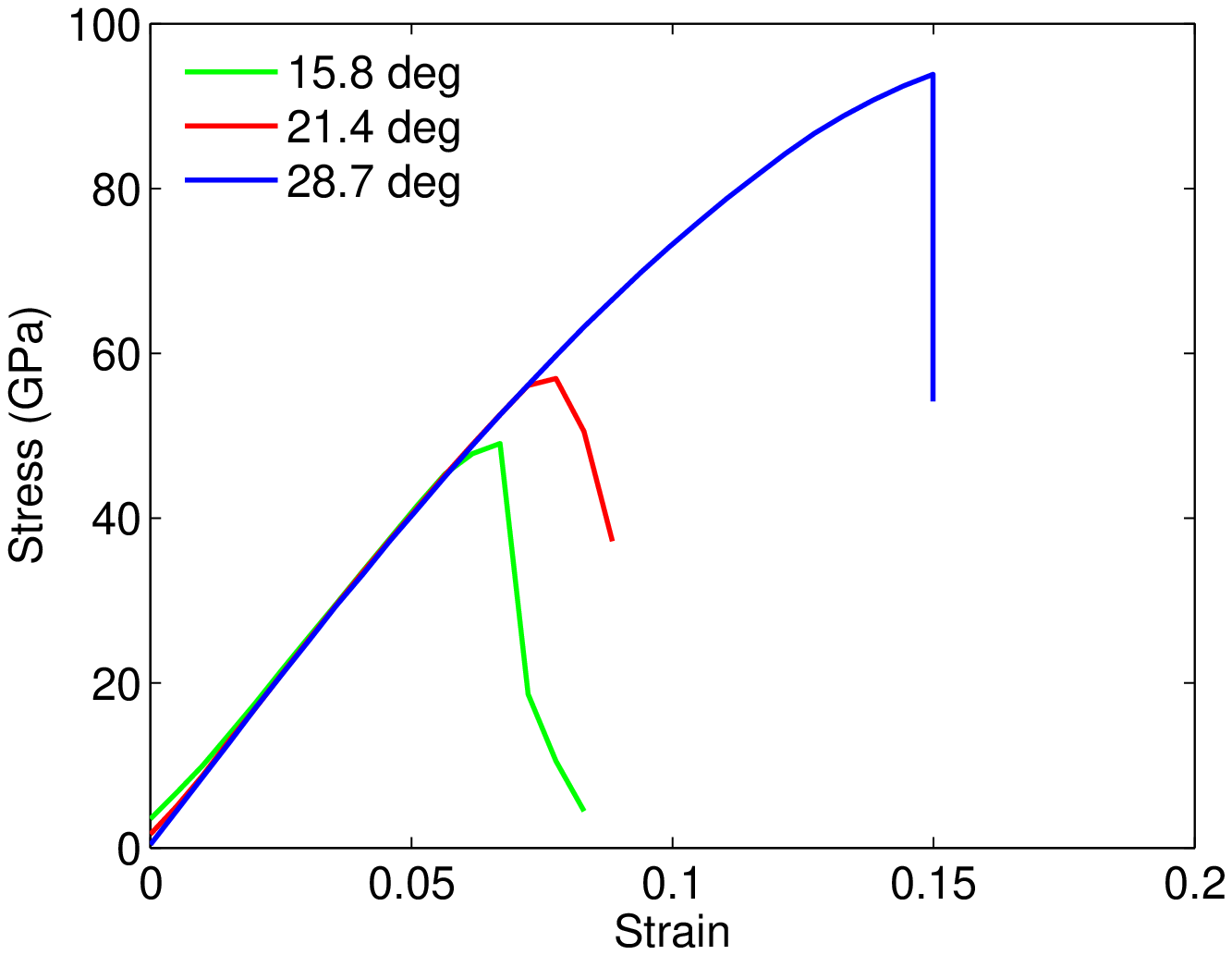}
    } \qquad
    \subfigure[Parallel]
    { \label(Parallel){fig4_sub2}
        \includegraphics[scale=0.5]{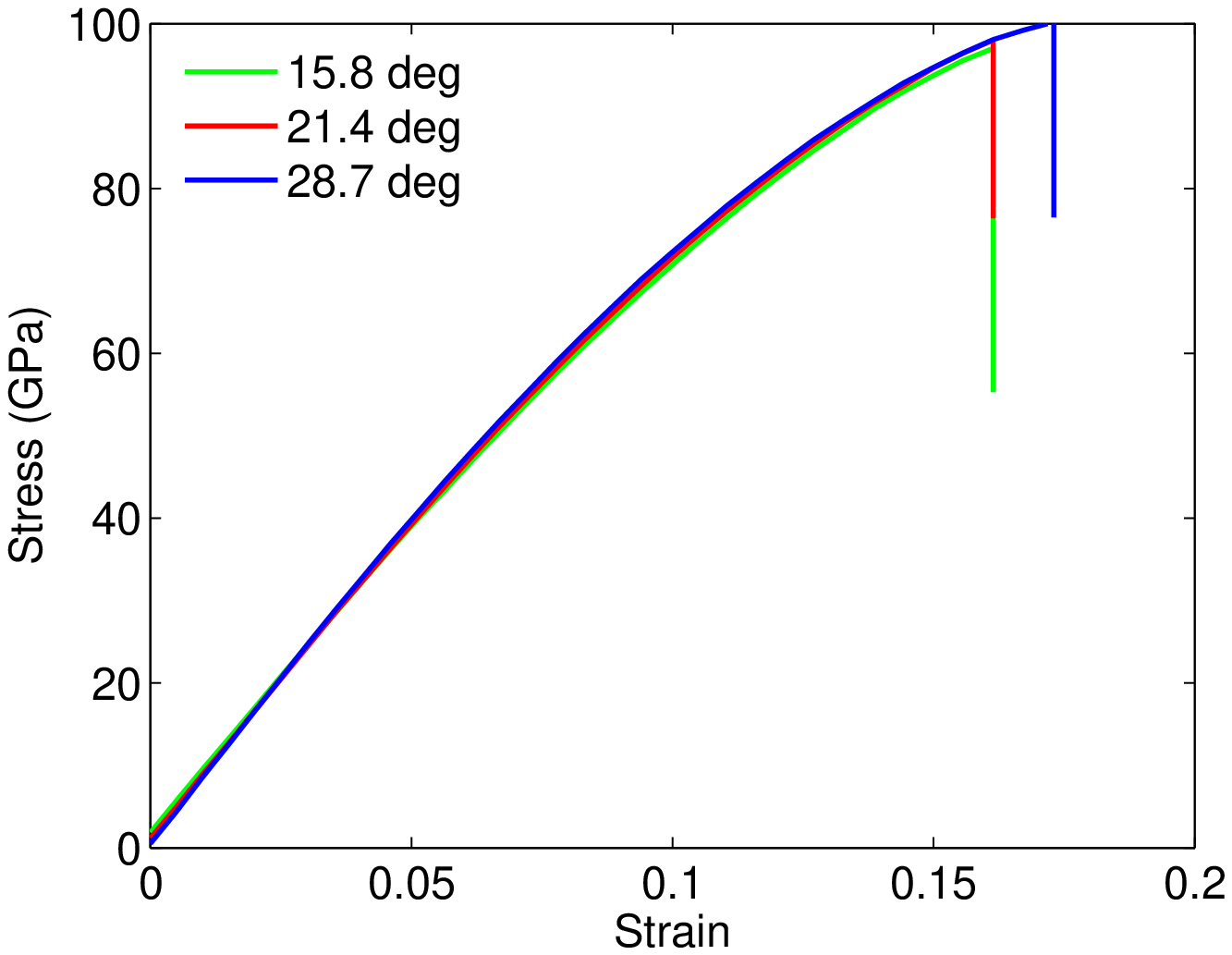} 
    } \\
    \caption{The stress-strain curves of arm-chair oriented graphene sheets pulled perpendicular (a) and parallel (b) to the grain boundaries.}
    \label{fig4}
\end{figure}

The stress-strain curves for zig-zag oriented graphene sheets with tilt grain boundaries are shown in Figure \ref{fig3}, while those of the arm-chair oriented graphene sheets are shown in Figure \ref{fig4}. For both orientations, the simulated stress-strain curves have been plotted for deformation perpendicular and parallel to the grain boundaries. In general, grain boundaries may be oriented at an angle relative to the tensile axes; to study the effect of this variation, we consider the extreme cases i.e., the grain boundaries oriented along and perpendicular to the loading axes. In both cases, the variation of the failure strength with angle is larger when the sheets are pulled perpendicular to the boundaries than when they are pulled parallel to the boundaries.

Upon initial inspection, the data in these plots looks mislabled, or it seems as though the legends have been misread, since all four plots illustrate the same completely counterintuitive result: \emph{As the grain boundary angle, and hence the defect density, increases, the ultimate failure strength and strain at failure increase!} One would automatically assume that as the number of defects increases, the strength of any material should \emph{decrease}; how is it that a higher density of 5-7 defects along the grain boundaries actually \emph{increases} the strength of graphene sheets?

To answer this question and to make an analogy to what continuum models predict about the scaling of strength with defect density, we consider a fracture-mechanics-based approach in which we model the 7-member rings along the grain boundary as an infinite array of Griffith cracks. This is a reasonable continuum-level analogue of the 7-member rings, since these rings are larger than the hex-rings and so they can be represented as a crack or a void within the material. This assumption is also consistent with the fact that failure in the graphene sheets \emph{always} begins at the 7-member rings (we will discuss this phenomenon in more detail later on in the paper). We therefore consider an infinite array of Griffith cracks (the crack tips aligned along the boundary), each of length ${2a}$, and separated by a length ${2h}$ from one-another. The common method that is used to determine whether a crack advances upon application of a remote stress, ${\sigma_{\infty}}$, is to compute the stress-intensity factor, ${K_I}$. If ${K_I}$ exceeds ${K_{IC}}$ - the experimentally measured fracture toughness for a given material - then crack propagation will ensue. The stress-intensity factor for the arrangement of cracks outlined above can be computed using standard fracture mechanics techniques \cite{broberg}, and is presented in the following equation:

\begin{equation}
K_I=\sigma_{\infty}\sqrt{(2h)tan\Big(\frac{\pi a}{2h}\Big)}
\label{eq6}
\end{equation}

\begin{figure}[!ht]
    \centering
    { 
       \includegraphics[scale=0.6]{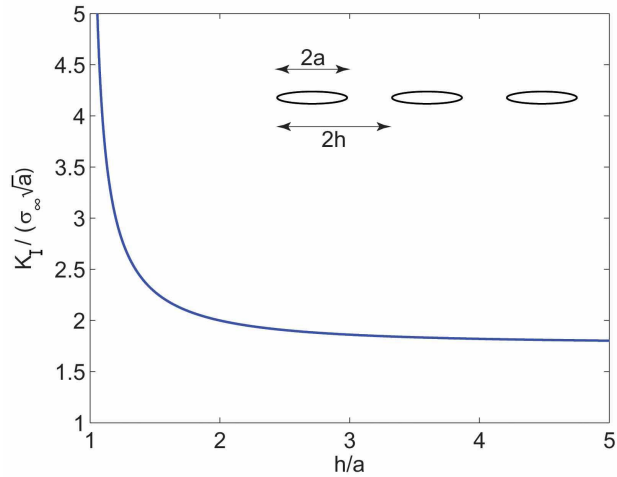}
    }
    \caption{Non-dimensional stress-intensity factor versus normalized crack-spacing.}
    \label{fig9}
\end{figure}

A plot of the non-dimensional stress-intensity factor, ${K_I/\sigma_{\infty}\sqrt{a}}$, versus normalized crack-spacing, ${h/a}$, is presented in Figure \ref{fig9}. As the inter-crack spacing, ${2h}$, decreases, the stress-intensity factor, ${K_I}$, increases due to the interaction of the stress fields of adjacent cracks. Based on the plot in Figure \ref{fig9}, graphene sheets \emph{should} be weaker as the defect distribution becomes more dense. Clearly, this fracture mechanics analogy fails to explain our results, and so the explanation that we seek does not lie within continuum mechanics techniques, but on the atomic-level details of bond rupture and failure. We therefore focus on the sequence of atomic-scale events that leads to tensile failure.

\begin{figure}[!ht]
    \centering
    \subfigure[${5.5^{\circ}}$]
    { \label(${5.5^{\circ}}$){fig5_sub1}
        \fbox{\includegraphics[scale=1.00]{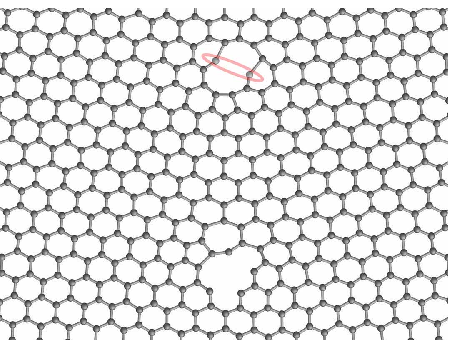}}
    } \qquad
    \subfigure[${13.2^{\circ}}$]
    { \label(${13.2^{\circ}}$){fig5_sub2}
        \fbox{\includegraphics[scale=1.00]{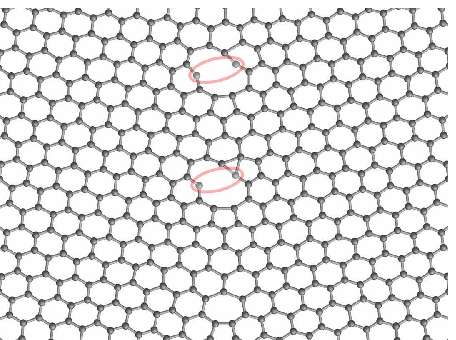}}
    } \qquad
    \subfigure[${21.7^{\circ}}$]
    { \label(${21.7^{\circ}}$){fig5_sub3}
        \fbox{\includegraphics[scale=1.00]{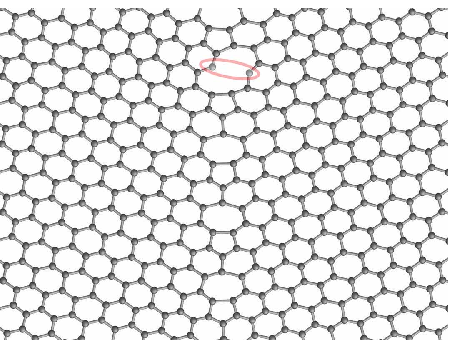}}
    } \\
    \caption{The initial stages of failure in zig-zag oriented graphene sheets pulled perpendicular to the grain boundaries.}
    \label{fig5}
\end{figure}

Figure \ref{fig5} shows the first signs of failure within the zigzag oriented graphene sheets during deformation perpendicular to the boundaries. What is striking is that the first bonds to break (the top-most bonds of the 7-rings on either the left or right side, highlighted in red) are always the same ones for all three grain boundary angles. Once these bonds have been broken, complete failure of the sheets proceeds rapidly along the grain boundaries.

\begin{figure}[!ht]
    \centering
    \subfigure[${5.5^{\circ}}$]
    { \label(${5.5^{\circ}}$){fig6_sub1}
        \fbox{\includegraphics[scale=1.00]{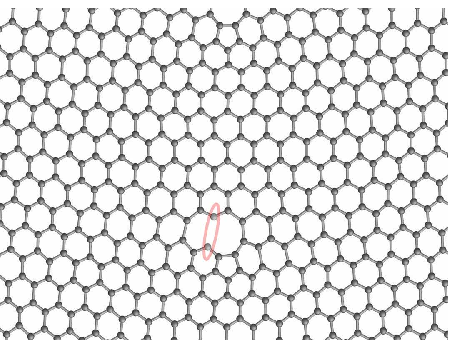}}
    } \qquad
    \subfigure[${13.2^{\circ}}$]
    { \label(${13.2^{\circ}}$){fig6_sub2}
        \fbox{\includegraphics[scale=1.00]{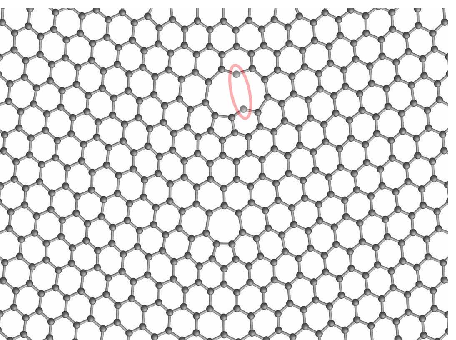}} 
    } \qquad
    \subfigure[${21.7^{\circ}}$]
    { \label(${21.7^{\circ}}$){fig6_sub3}
        \fbox{\includegraphics[scale=1.00]{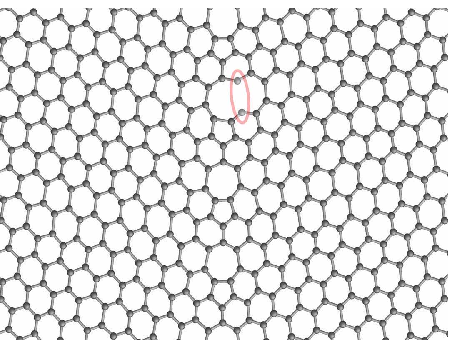}}
    } \\
    \caption{The initial stages of failure in zig-zag oriented graphene sheets pulled parallel to the grain boundaries.}
    \label{fig6}
\end{figure}

The first signs of failure for the zig-zag oriented graphene sheets pulled parallel to the grain boundaries are shown in Figure \ref{fig6}. As was the case previously, we observe that for each of the three grain boundary angles, the same bonds in the 7-rings (the most vertically aligned side bonds in this case, highlighted in red) are the first to break, although it should be noted that the critical bonds in this case are different from those of the sheets that were pulled perpendicular to the boundary. The location of the critical bonds is dependent on the orientation of the graphene (zig-zag or arm-chair) and the loading direction (parallel or perpendicular to the grain boundary); for each specific combination of orientation and loading direction, the critical bonds are the same for all three grain boundary angles.

\begin{figure}[!ht]
    \centering
    \subfigure[${15.8^{\circ}}$]
    { \label(${15.8^{\circ}}$){fig7_sub1}
        \fbox{\includegraphics[scale=1.00]{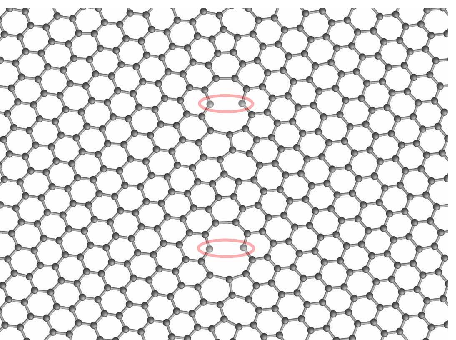}}
    } \qquad
    \subfigure[${21.4^{\circ}}$]
    { \label(${21.4^{\circ}}$){fig7_sub2}
        \fbox{\includegraphics[scale=1.00]{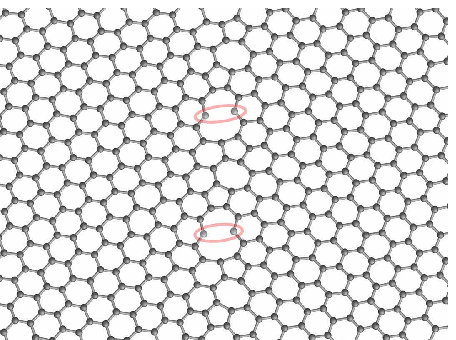}}
    } \qquad
    \subfigure[${28.7^{\circ}}$]
    { \label(${28.7^{\circ}}$){fig7_sub3}
        \fbox{\includegraphics[scale=1.00]{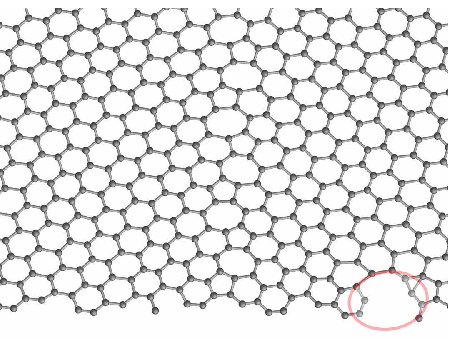}}
    } \\
    \caption{The initial stages of failure in arm-chair oriented graphene sheets pulled perpendicular to the grain boundaries.}
    \label{fig7}
\end{figure}

\begin{figure}[!ht]
    \centering
    \subfigure[${15.8^{\circ}}$]
    { \label(${15.8^{\circ}}$){fig8_sub1}
        \fbox{\includegraphics[scale=1.00]{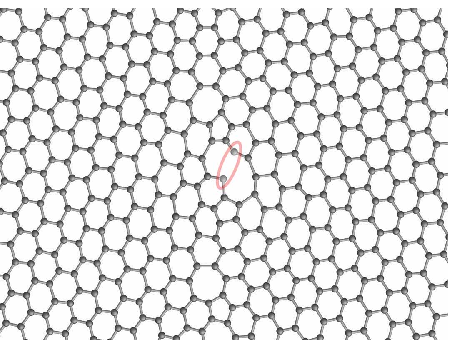}}
    } \qquad
    \subfigure[${21.4^{\circ}}$]
    { \label(${21.4^{\circ}}$){fig8_sub2}
        \fbox{\includegraphics[scale=1.00]{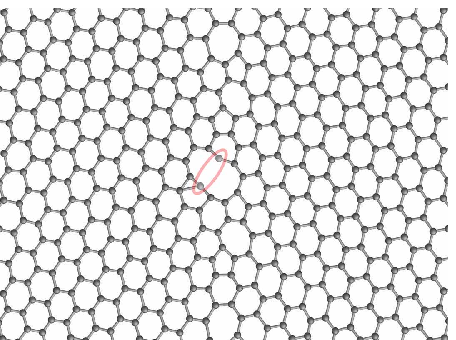}}
    } \qquad
    \subfigure[${28.7^{\circ}}$]
    { \label(${28.7^{\circ}}$){fig8_sub3}
        \fbox{\includegraphics[scale=1.00]{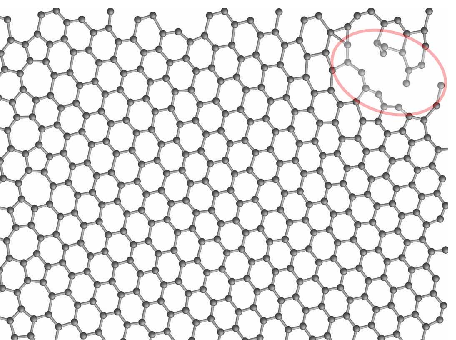}}
    } \\
    \caption{The initial stages of failure in arm-chair oriented graphene sheets pulled parallel to the grain boundaries.}
    \label{fig8}
\end{figure}

We now focus our attention on the incipient failure of arm-chair oriented graphene sheets pulled perpendicular and parallel to the boundary, as shown in Figures \ref{fig7} and \ref{fig8}, respectively. As was the case for zig-zag oriented graphene, we observe that for each loading direction, the same critical bonds (those highlighted in red) initiate failure in the arm-chair oriented graphene sheets regardless of the grain boundary angle. There is, however, an exception in the case of the largest grain boundary angle for both loading directions, which actually fail away from the boundaries, within the highlighted regions in Figures \ref{fig7_sub3} and \ref{fig8_sub3}.

Having identified the critical bonds, we now focus on the initial strains in these bonds as a function of the grain boundary angle, and uncover clues towards understanding the anomalous strength of tilt grain boundaries in graphene. As the grain boundary angle increases, the initial lengths of the critical bonds decrease towards the ${sp^2}$ carbon-carbon bond-length in pristine graphene. Prior to any applied deformation, for loading perpendicular to the boundary, the strain in the critical bonds of the zigzag oriented graphene sheets with grain boundary angles of ${5.5^{\circ}}$, ${13.2^{\circ}}$, and ${21.7^{\circ}}$, are ${12.2\%}$, ${10.3\%}$, and ${5.4\%}$, respectively. Our DFT simulations validate these results and the general trend, with calculated strains of ${9.5\%}$, ${8.7\%}$, and ${5.4\%}$ as the grain boundary angle increases. Naturally, as the pre-strain in the material decreases, the strain at failure and ultimate strength will increase. It is the level of preexisting strain within the critical bonds of the 7-member rings that accounts for the counterintuitive results we have observed in our simulations.

The initial strains in the critical bonds for zig-zag graphene pulled parallel to the boundary are ${2.1\%}$, ${1.7\%}$, and ${0.7\%}$ for the ${5.5^{\circ}}$, ${13.2^{\circ}}$, and ${21.7^{\circ}}$ grain boundary angles, respectively. The strains calculated through DFT are slightly higher at ${3.2\%}$, ${2.2\%}$, and ${1.7\%}$, however, the trend matches that of the MD simulations perfectly. Although the level of initial strain is lower for these critical bonds than those discussed in the preceding paragraph, the general trend of decreasing strain with increasing grain boundary angle is the same, and is consistent with the stress-strain results plotted in Figure \ref{fig3}.

In the undeformed state, the critical bonds in arm-chair graphene pulled perpendicular to the boundary are strained by a factor of ${23.4\%}$, ${9.3\%}$, and ${1.7\%}$ for the ${15.8^{\circ}}$, ${21.4^{\circ}}$, and ${28.7^{\circ}}$ boundary angles, respectively. Once again, we observe the trend of decreasing initial strain with increasing grain boundary angle. Interestingly, the graphene sheet with a ${28.7^{\circ}}$ grain boundary angle begins to fail away from the boundary, at the location highlighted in Figure \ref{fig7_sub3}. This is due to the fact that in this case, the bond lengths in the 7-member rings are very close to those of pure graphene (the previously mentioned strain of ${1.7\%}$ being the largest among the 7 bonds), and two of the bonds are actually initially shorter than those of pure graphene. Reexamination of the stress-strain curve corresponding to this grain boundary angle and pulling direction (shown in Figure \ref{fig4_sub1}) indicates a strain at failure of ${15.5\%}$, and an ultimate strength of ${95GPa}$, values that are approaching the strength of pure arm-chair graphene. Based on these results, we can conclude that grain boundaries with a mismatch angle of ${28.7^{\circ}}$ do not affect the strength of  arm-chair oriented graphene sheets appreciably, whereas those with lower separation angles weaken them significantly.

For arm-chair graphene pulled parallel to the boundary, the critical bonds in the ${15.8^{\circ}}$ and ${21.4^{\circ}}$ are strained by factors of ${5.4\%}$ and ${4.0\%}$, respectively; the ${28.7^{\circ}}$ sheets begin to fail away from the boundary (within the highlighted region in Figure \ref{fig8_sub3}), due to the fact that for the most severe grain boundary angle, this bond is actually the same length as those in pure graphene. Thus, we have shown that the general trend of decreasing initial strain with increasing grain boundary angle is perfectly consistent for zigzag and arm-chair oriented graphene sheets, and that the initial bond lengths fully explain the counterintuitive results observed in our MD simulations.


In summary, we used MD and DFT calculations to study the mechanical strength of grain boundaries in zig-zag and arm-chair oriented graphene sheets. For both orientations, we have found that the strain at failure and ultimate strength of graphene increases with grain boundary angle. We have looked in detail at the atomic-scale bond-breaking processes that lead to failure and have identified the critical bonds that determine the ultimate strength of the grain boundaries. Based on these analyses, it is clear that the initial strain in these bonds determines the failure strength - the higher the strain, the lower the strength. Higher grain boundary angles can better accommodate the 7-ring defects that comprise the grain boundaries, therefore the initial strain in the critical bonds decreases with increasing angle. Fracture mechanics methods were unable to predict the trends from our simulations because the influence of strained atomic bonds is inherently absent from continuum techniques. 


%

\end{document}